\documentclass[letterpaper, 10 pt, conference]{ieeeconf}  

\IEEEoverridecommandlockouts                              

\usepackage{soul, xcolor}
\usepackage{comment}
\usepackage{multirow}
 \usepackage{graphicx}
 \usepackage{subcaption}
 \usepackage{multirow}
 \usepackage{amsmath,amssymb,amsfonts}
 \usepackage{wrapfig}
\makeatletter
\newcommand\notsotiny{\@setfontsize\notsotiny\@vipt\@viipt}
\makeatother

\title{\LARGE \bf
Frame-Level Real-Time Assessment of Stroke Rehabilitation Exercises from Video-Level Labeled Data: Task-Specific vs. Foundation Models
}

\author{Gonçalo Mesquita$^{1}$, Ana Rita Cóias$^{1}$ \IEEEmembership{Student Member, IEEE}, Artur Dubrawski$^{2}$ \IEEEmembership{Member, IEEE}, \\ and Alexandre Bernardino$^{1}$ \IEEEmembership{Senior Member, IEEE}
\thanks{This work was supported by the Portuguese Foundation for Science and Technology - FCT by LARSyS FCT funding (DOI: 10.54499/LA/P/0083/2020, 10.54499/UIDP/50009/2020, and 10.54499/UIDB/50009/2020), FCT HAVATAR Project (DOI: 10.54499/PTDC/EEI-ROB/1155/2020), by FCT under CMU Portugal, by the Portuguese Recovery and Resilience Plan (RRP), through project number 62, Center for Responsible AI, by NSF awards 2427948 and 2406231, and by DARPA award HR00112420329. Ana Rita Cóias is supported by the FCT doctoral grant [SFRH/BD/05239/2021].}
\thanks{$^{1}$Institute for Systems and Robotics, Laboratory for Robotics and Engineering Systems, Instituto Superior Técnico, University of Lisbon, Lisbon, Portugal
        {\tt\small goncalo.mesquita@tecnico.ulisboa.pt; ana.coias@tecnico.ulisboa.pt; alex@isr.tecnico.ulisboa.pt}}%
\thanks{$^{2}$Auton Lab, School of Computer Science, Carnegie Mellon University, Pittsburgh, PA, USA
        {\tt\small awd@cs.cmu.edu}}%
}

\usepackage[switch]{lineno}
\usepackage{float}
\begin{document}

\maketitle
\thispagestyle{empty}
\pagestyle{empty}


\begin{abstract}
The growing demands of stroke rehabilitation have increased the need for solutions to support autonomous exercising. Virtual coaches can provide real-time exercise feedback from video data, helping patients improve motor function and keep engagement. However, training real-time motion analysis systems demands frame-level annotations, which are time-consuming and costly to obtain. In this work, we present a framework that learns to classify individual frames from video-level annotations for real-time assessment of compensatory motions in rehabilitation exercises. We use a gradient-based technique and a pseudo-label selection method to create frame-level pseudo-labels for training a frame-level classifier. We leverage pre-trained task-specific models - Action Transformer, SkateFormer - and a foundation model - MOMENT - for pseudo-label generation, aiming to improve generalization to new patients. To validate the approach, we use the \textit{SERE} dataset with 18 post-stroke patients performing five rehabilitation exercises annotated on compensatory motions. MOMENT achieves better video-level assessment results (AUC = $73\%$), outperforming the baseline LSTM (AUC = $58\%$). The Action Transformer, with the Integrated Gradient technique, leads to better outcomes (AUC = $72\%$) for frame-level assessment, outperforming the baseline trained with ground truth frame-level labeling (AUC = $69\%$). We show that our proposed approach with pre-trained models enhances model generalization ability and facilitates the customization to new patients, reducing the demands of data labeling.
\newline

\indent \textit{Index Terms}— Compensatory Motion Patterns, Stroke Rehabilitation, Real-time Motion Assessment, Saliency Maps, Weakly Supervised Learning, Transfer Learning, Transformers, Tasks-Specific Models, Foundation Model.

\end{abstract}

\section{INTRODUCTION}

After a stroke, prompt rehabilitation therapy with task-oriented exercises is crucial for motor function recovery \cite{rensink2009task,schneider2019extra,billinger2014physical}. Therapists evaluate motor function, guide exercises, and provide proper feedback \cite{serrada2016current}. With the increasing patient load, high rehabilitation costs, and therapist shortages \cite{meadmore2019factors, watkins2023challenges}, recommendations for autonomous exercise at home or between therapy sessions have grown  \cite{peek2016interventions}. Exercising alone is notably challenging as the lack of guidance and feedback highly impacts motivation and engagement, hampering recovery \cite{meadmore2019factors,maclean2000qualitative}. This has drawn interest in rehabilitation support solutions research, such as Virtual Coaches (VCs). VCs must assess exercise performance and provide real-time feedback helping to improve motor function \cite{siewiorek2012architecture,weimann2022virtual}.  

Advances in computer vision and machine learning have made it possible to automatically and objectively assess movement ability from videos \cite{ozturk2016clinically, olesh2014automated, lee2019learning}. While reviewing performance after exercises needs video-level labels \cite{lee2019learning}, giving feedback during exercises requires detailed labels at the clip or frame level. However, collecting fully labeled datasets is expensive, time-consuming, and often impractical in many real-world situations \cite{lanotte2023ai, parnami2203learning}.

 Past research mainly used fully supervised models to provide real-time (frame-level) feedback on stroke rehabilitation exercises \cite{lee2020towards, coias2022low, mennella2023deep}. Lee \textit{et al.} \cite{lee2023exploring} used a gradient-based technique to generate frame-level pseudo-labels from salient features and frames, identifying compensatory patterns (e.g., shoulder elevation) in video frames from a full motion trial assessment. However, further evaluation on pseudo-labels usability for real-time assessment was missing. Later, Cóias et al. \cite{Coias2025} introduced a framework using saliency maps and a threshold method to create pseudo-labels from salient kinematic measures and frames. These labels trained a model to detect incorrect movements at the frame level, aiming to work even on weakly labeled datasets.
In semantic segmentation, Class Activation Mapping (CAM) for pixel-level label assignment has been used for training semantic segmentation modes, enhancing their performance \cite{chen2023segment}. In action recognition, video clip-level pseudo-labels have been generated by leveraging attention scores to facilitate action localization \cite{yu2023frame,xu2022cross}. 

To effectively support rehabilitation exercises or clinical decision-making, solutions should ensure robust model generalization to new patients as a single solution must accommodate a diverse population. However, it is impractical to train models from scratch for each new patient \cite{hosna2022transfer}. Using transfer learning, such as leveraging pre-trained models on large datasets, enhances generalization when the source and target domains or tasks are identical or somehow related \cite{hosna2022transfer}. These techniques have proven successful in human action recognition tasks \cite{abdulazeem2021human,wang2017internal, kilicc2024skelresnet} with small datasets.

In this work, we introduce a framework that classifies individual frames using video-level annotations for real-time assessment of compensatory motions in rehabilitation exercises. Our approach leverages pre-trained task-specific models — Action Transformer (AcT) \cite{mazzia2022action} and SkateFormer \cite{do2025skateformer} — trained on large action recognition datasets, and the time-series foundation model MOMENT \cite{goswami2024moment}, which performance we compare against a Long Short-Term Memory (LSTM) network baseline for video-level assessment. We employ a gradient-based technique and a pseudo-label selection method to generate frame-level pseudo-labels from video-level predictions, which we use to train a frame-level classifier. We explore Vanilla Gradient and Integrated Gradient techniques and compare different strategies for pseudo-label selection. By using pre-trained models, we aim to enhance video-level generalization outcomes for new patients, improving quality in pseudo-labeling and subsequent frame-level classification. We evaluate our approach using the \textit{SERE} dataset, which includes 18 post-stroke patients performing five functional exercises annotated on compensatory motions.

MOMENT achieves superior video-level assessment (AUC = $73\%$), surpassing the baseline LSTM (AUC = $58\%$). For frame-level assessment, AcT delivers the best results using Integrated Gradients for pseudo-label generation (AUC = $72\%$) outperforming the baseline trained with ground-truth frame-level labels (AUC = $69\%$). Frame-level ground-truth labels are only used for training the baseline frame-level classifier and for validation purposes. Our approach, leveraging pre-trained models, enhances generalization and simplifies customization to new patients, reducing data labeling efforts.

This work provides the following contributions:

\begin{itemize}
    \item We explore the advantage of using pre-trained task-specific models, trained on action recognition datasets, and a foundation model for time-series on generalization outcomes to new patients on video-level compensatory motion assessment in stroke rehabilitation exercises;
    \item We investigate the impact of improved video-level assessment on frame-level pseudo-label generation by evaluating subsequent frame-level classification results across models used in the pipeline;
    \item We explore Vanilla and Integrated gradient techniques jointly with a single or dual threshold pseudo-label selection for frame-level pseudo-label selection.
\end{itemize}




\section{METHODS}

\subsection{Problem Definition}

\begin{wrapfigure}{r}{0.24\textwidth}
    \centering
      \includegraphics[scale=0.27]{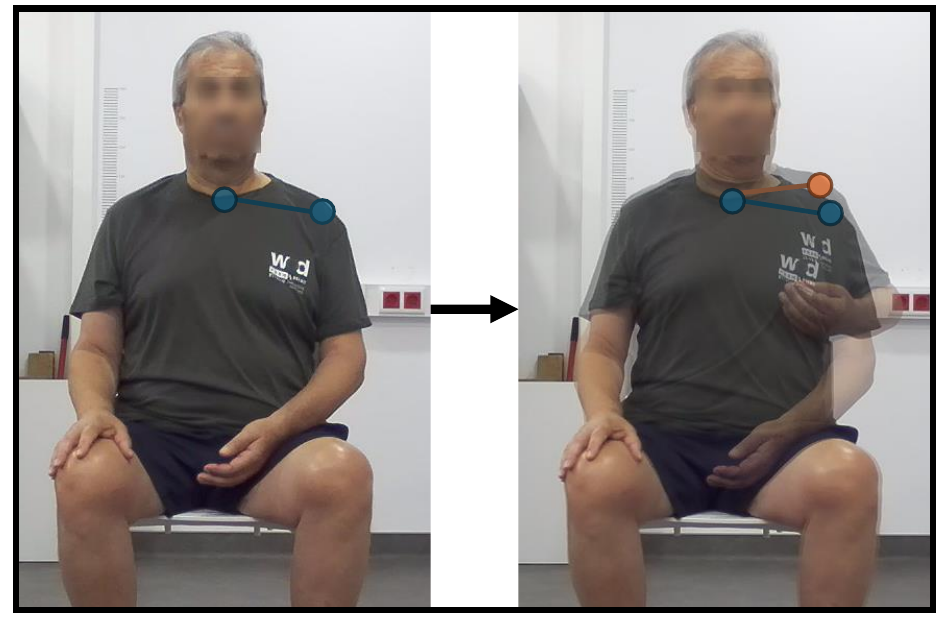}
      \caption{Example of a shoulder elevation motion.}
      \label{fig:compex}
      \vspace{-0.2cm}
\end{wrapfigure}

We consider a set of N untrimmed videos of post-stroke patients performing a functional exercise motion trial, $V = \{v^i\}^N_{i=1}$, $i$ denotes the video number. Each video has a label, $y^i = \{0,1\}$, denoting the existence of compensation in the described motion (0: normal, 1: compensatory motion). Post-stroke patients may describe compensatory motions — new movement patterns adopted to enable task completion (e.g., reaching an object) \cite{levin2009motor}. Excessive trunk flexion, shoulder elevatio, and head flexion are common compensatory motions. Figure \ref{fig:compex} illustrates shoulder elevation compensatory motion.

Using a gradient-based technique, we generate a saliency map to highlight video frames where compensatory motions occur. The gradients are used for pseudo-label creation. In this paper, we denote as $f^i$ the video features, $z^i_t$ the pseudo-label for the frame $t$ of the video $i$, and $T$ is the maximum number of frames. 


\subsection{Approach Pipeline Overview}

\begin{figure}[thpb]
      \centering
      \includegraphics[scale=0.48]{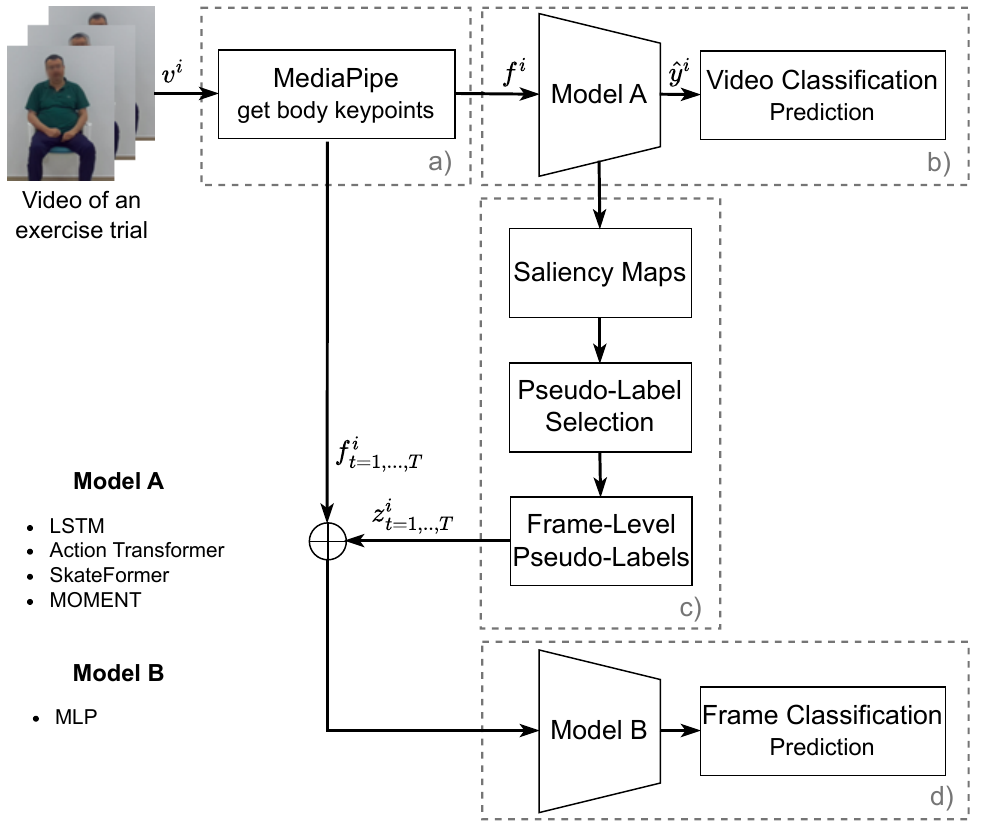}
      \caption{Approach Pipeline: a) body pose extraction and preprocessing; b) video-level assessment with Model A; c) if a compensatory motion is detected, through the gradient-based technique we generate a saliency map to which we apply a pseudo-labels selection method to generate frame-level pseudo-labels; d) training of Model B, a Multilayer Perceptron (MLP), for frame-level assessment comparing the ground truth with the created frame-level pseudo-labels.}
      \label{fig:pipeline}

   \end{figure}

Figure \ref{fig:pipeline} illustrates our approach pipeline for real-time video assessment. We consider as features of analysis body pose keypoints extracted by state-of-the-art human body pose detectors (box a) in Figure \ref{fig:pipeline}). We preprocess the extracted data to reduce noise artifacts (box a) in Figure \ref{fig:pipeline}). From the preprocessed features and video-level labels we fine-tune a video classifier (Model A) - AcT  \cite{mazzia2022action}, SkateFormer \cite{do2025skateformer}, MOMENT \cite{goswami2024moment}, or LSTM - for video-level compensatory motion assessment (box b) in Figure \ref{fig:pipeline}). From Model A predictions, we apply a gradient-based technique to generate saliency maps \cite{simonyan2013deep,sundararajan2017axiomatic}. These maps are subsequently used to produce frame-level pseudo-labels by selecting the most salient features and frames using a pseudo-label selection method (box c) in Figure \ref{fig:pipeline}). Finally, with the same training data and generated pseudo-labels, we train a Multilayer Perceptron (MLP) for frame-level compensation classification (box d) in Figure \ref{fig:pipeline}). 

\newpage
\subsection{Pose Estimation and Preprocessing}

\begin{wrapfigure}{r}{0.2\textwidth}
\vspace{-0.55cm}
    \centering
      \includegraphics[scale=0.13]{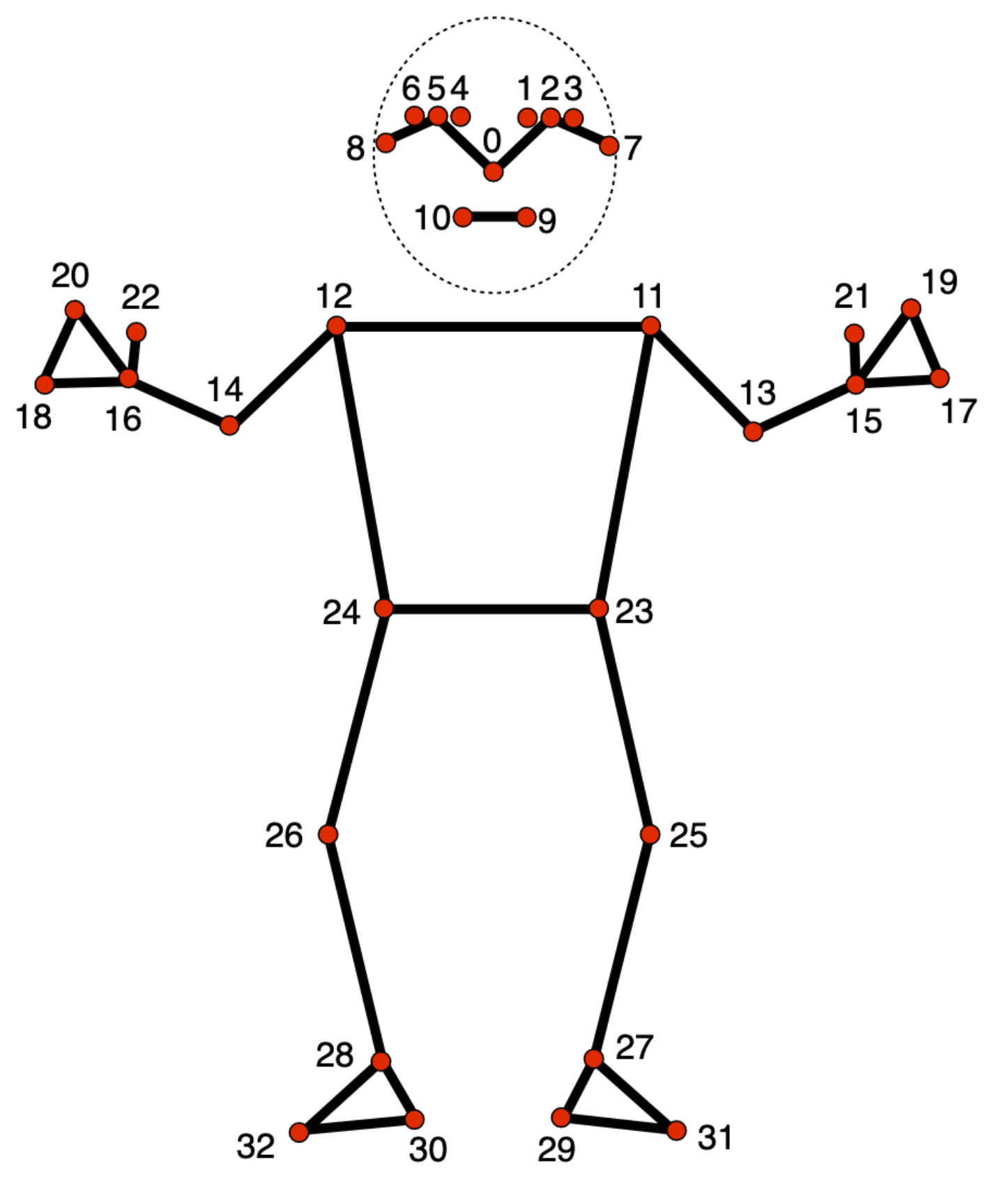}
      \caption{ MediaPipe Pose Keypoints.}
      \label{fig:skeleton}
      \vspace{-0.25cm}
\end{wrapfigure}

The pose estimation and preprocessing steps extract key motion features, reduce noise, and enhance model generalization. 
MediaPipe\footnote{https://ai.google.dev/edge/mediapipe/solutions/vision/pose\_landmarker} is used to track post-stroke patients' 
movements, providing 33 joint keypoints per frame in real-world 3D coordinates. Figure \ref{fig:skeleton} shows the extracted keypoints.
Its efficiency, robustness to occlusions, and real-time capabilities make it well-suited for stroke rehabilitation  \cite{coias2023skeleton}.


Joint positions in each frame were offset by their initial trial positions, highlighting movement changes over absolute positions. This approach reduces starting position variability and improves the detection of subtle compensatory motions by representing motion trajectories as displacement vectors. We applied a moving average filter with a window of five frames to smooth small variations in the extracted signal.

\subsection{Video-Level Compensation Assessment - Task Specific vs. Foundation Models}

This study compares task-specific models, trained on action recognition datasets, Action Transformer (AcT) \cite{mazzia2022action} and SkateFormer \cite{do2025skateformer}, with the MOMENT \cite{goswami2024moment} time-series foundation model, pre-trained models on large and diverse datasets to tackle multiple tasks. We fine-tuned these models with a rehabilitation exercises dataset. We aim to identify which model yields better video-level assessment outcomes and subsequently improves frame-level assessment results regarding the generalization to new patients by enhancing the quality of pseudo-labels. 

The AcT, a fully self-attention based architecture inspired by the Vision Transformer \cite{dosovitskiy2021thomas}, is tailored for short-time, pose-based human action recognition. Its self-attention mechanisms capture temporal dependencies, focusing on key frames where compensatory movements are most pronounced. AcT was pre-trained and evaluated on the MPOSE2021 dataset \cite{mazzia2022action}, a large-scale comprehensive collection of human pose data derived from OpenPose and PoseNet across multiple human action recognition datasets. 

Likewise, SkateFormer \cite{do2025skateformer} is a skeletal-temporal transformer model designed for skeleton-based action recognition. It uses a partition-specific attention mechanism, classifying skeletal-temporal relationships into four categories based on joint proximity and temporal adjacency. This strategy enables SkateFormer to prioritize key joints and frames for action recognition, effectively capturing complex compensatory movements that can differ across tasks. SkateFormer was pre-trained on the NTU RGB+D 120 dataset \cite{liu2019ntu}, which offers a comprehensive collection of skeleton sequences for diverse human actions. Its effectiveness in capturing skeletal-temporal relationships has been demonstrated through evaluations on datasets such as NTU RGB+D 60 \cite{shahroudy2016ntu}, NTU RGB+D 120 \cite{liu2019ntu}, and NW-UCLA \cite{wang2014cross}.

Unlike task-specific models, MOMENT \cite{goswami2024moment} is pre-trained on diverse time-series tasks such as classification, forecasting, anomaly detection, and imputation, using large-scale datasets from various time-series domains. This extensive pre-training equips MOMENT with a general understanding of time-series dynamics, enabling rapid adaptation to new scenarios, including the classification of compensatory movements in stroke rehabilitation.

The hypothesis is that pre-trained models enhance generalization, reducing the need for extensive data collection and retraining. Fine-tuning foundation models for rehabilitation tasks can provide reliable patient-specific predictions, even with limited data or varying patient characteristics.

For video-level compensatory motion assessment, an LSTM exclusively trained on the rehabilitation dataset serves as the baseline, configured as a Many-to-One model with a single layer and a hidden size of 192. The AcT, SkateFormer, and MOMENT models retain their original architectures.

\subsection{Frame-Level Pseudo-Labels Generation} 

Following \cite{Coias2025}, we generate frame-level pseudo-labels from saliency maps highlighting significant features and frames from the video-level predictions. To create the saliency maps, we explore two gradient-based techniques: the Vanilla Gradient and the Integrated Gradient.




The Vanilla Gradient \cite{simonyan_deep_2014} method computes the gradient of the model’s output with respect to input features, identifying areas where small input changes significantly impact predictions. The saliency map is derived from the absolute gradient values, highlighting influential features.  

The Integrated Gradients \cite{sundararajan_axiomatic_2017}  method extends Vanilla Gradient by integrating gradients along a path from a baseline input to the actual input, providing a more comprehensive feature attribution. Both methods generate saliency maps that emphasize key frames relevant to compensatory motion detection, even for unseen patients.

\subsubsection{Pseudo-label Selection Method}  

From the saliency maps, we aggregate gradients frame by frame and apply min-max normalization to scale the results to a range of \([0,1]\), yielding a pseudo-score \(s^i_t\) for each frame \(t\). We distinguish the frames of a normal motion from the frames of a compensatory motion by thresholding frames pseudo-scores. We explore single threshold and dual threshold approaches.

\noindent \textbf{Single threshold:}
This approach requires only one threshold \(\tau\).  Using a threshold, $\tau$, each frame is assigned with a pseudo-label, $z^i_t$, \vspace{-0.2cm}

\begin{equation}
\small
    z^i_t = 
    \begin{cases}
    0, \text{ if } \hat{y}^i = 0 \\
    \mathbb{I}(s^i_t > \tau), \text{ if } \hat{y}^i = 1
    \end{cases}
\end{equation}

\noindent where $\hat{y}^i$ represents the predicted class for video $i$, and $\mathbb{I}$ is the indicator function. For normal motion video trial ($\hat{y}^i = 0$), all frames are assigned with a pseudo-label $z^i_t = 0$. For videos with compensatory motions ($\hat{y}^i = 1$), each frame's pseudo-score $s^i_t$ is compared against the threshold $\tau$. If $s^i_t > \tau$, the indicator function assigns a frame pseudo-label $z^i_t = 1$ and $z^i_t = 0$, otherwise.

\noindent \textbf{Dual threshold:} This approach uses two thresholds \(\tau_1\) and \(\tau_2\), with \(\tau_1 < \tau_2\). Therefore each frame is assigned with a pseudo-label, $z^i_t$, 

\begin{equation}
\small
    z^i_t = 
    \begin{cases}
    0,  & \text{if } \hat{y}^i = 0 \vee (s^i_t < \tau_1 \wedge \hat{y}^i = 1) \\
    1, & \text{if } s^i_t > \tau_2 \wedge \hat{y}^i = 1\\
    \text{Not Used}, & \text{if } \tau_1 \le s^i_t \le \tau_2 \wedge \hat{y}^i = 1.
    \end{cases}
\end{equation}

\noindent where, frames with scores below \(\tau_1\) are confidently considered normal (labeled 0), while those above \(\tau_2\) are confidently considered compensatory (labeled 1). The region \(\tau_1 \le s^i_t \le \tau_2\) may be treated as uncertain, therefore those labels are excluded and not used. 
This two-threshold strategy can reduce noisy frame-level assignments by clearly separating highly likely normal or compensatory frames from uncertain ones, however also has less data to feed Model B.

\subsection{Frame-Level Compensation Assessment}

Using the training set and the generated pseudo-labels, we train a fully supervised classifier for compensatory motion detection at the frame level, reducing reliance on costly data labeling. In Model B, we train a Multilayer Perceptron (MLP) and explore different model architectures, testing one to two layers with varying hidden units (32,48,64,96,128,192,256) for a binary classification task.

\section{EXPERIMENTS}

\subsection{StrokE Rehab Exercises (\textit{SERE}) dataset} 

\textit{SERE} \cite{Coias2025} is a newly collected dataset of 18 post-stroke patients performing five rehabilitation exercises. Table \ref{tab:exercises} details the exercises. In exercise 1 (E1), a patient raises their arm toward the head, simulating the action of combing their hair. In exercise 2 (E2), a patient has to move affected or unaffected arms towards the mouth and move it like brushing the teeth. In exercise 3 (E3), a patient must move both arms, simulating washing the face. In exercise 4 (E4), a patient has to tilt the trunk and move their arms towards the foot as if putting on socks for both feet. In exercise 5 (E5), a subject must lift their knee, flexing the hip with each leg.

Post-stroke patients performed ten repetitions of each exercise, with affected or unaffected arm\footnote{After a stroke, patients often describe weakness or loss of movement in one body side (hemiparesis).} (E1 and E2), both arms simultaneously (E3 and E4), and both legs (E5). 
The dataset has 1260 videos in which 538 compensatory motions are displayed (E1: 170, E2: 118, E3: 100, E4: 30, E5: 120).  

\begin{table}[ht]
\caption{ Functional exercises for stroke rehabilitation and corresponding joint motions.\vspace{-0.2cm}}
\label{tab:exercises}
\begin{center}
\scriptsize
\begin{tabular}{|c||c||l|}
\hline
\textbf{Exercise} & \textbf{Description} & \multicolumn{1}{c|}{\textbf{Motions}} \\
\hline
E1 & \textit{`Brushing Hair'} & \begin{tabular}[c]{@{}l@{}} $\bullet$ Shoulder flexion and elbow flexion/extension \end{tabular} \\ \hline
E2 & \textit{`Brushing Teeth'} & \begin{tabular}[c]{@{}l@{}} $\bullet$ Shoulder flexion and horizontal \\ abduction/adduction and elbow flexion/extension \end{tabular} \\ \hline
E3 & \textit{`Wash the Face'} &  \begin{tabular}[c]{@{}l@{}} $\bullet$ Elbow flexion, shoulder flexion/extension\\and abduction/adduction, and arm coordination \end{tabular}  \\ \hline
E4 & \textit{`Put on Socks'} & \begin{tabular}[c]{@{}l@{}} $\bullet$ Trunk flexion and slight right/left rotation,\\ shoulder flexion and elbow flexion/extension \end{tabular} \\ \hline
E5 & \textit{`Hip Flexion'} & \begin{tabular}[c]{@{}l@{}} $\bullet$ Hip flexion \end{tabular} \\ 
\hline
\end{tabular}
\end{center}
\vspace{-0.6cm}
\end{table}

\subsubsection{Data Collection}
The videos were recorded at a frame rate of 30 fps with a ZED Mini Stereo Camera from StereoLabs\footnote{https://www.stereolabs.com/} and the ZED Explorer framework from the ZED SDK.
The camera was positioned 2.5 meters from the patients and 0.90 meters above the floor. To ensure their safety at all times, patients performed the exercises while seated in a chair. Data collection occurred at the NeuroSer Rehabilitation Center and the Alcoitão Rehabilitation Medicine Center. 

Data collection complies with the General Data Protection Regulation (GDPR) and was approved by NeuroSer and Alcoitão Rehabilitation Center ethics committees.


\subsubsection{Participants}

Data collection involved 18 post-stroke patients (6 females and 12 males), with $61.22 \pm 15.06$ years old, $12.73 \pm 32.49$ months after the stroke, whose profiles are detailed in \cite{Coias2025}. 
Due to ethical reasons and consents, P02 and P04 \cite{Coias2025} where excluded from this study.  

\subsubsection{Annotation}

Two physiotherapists and a occupational therapist, with 9.33 $\pm$ 1.25 yeas of experience in stroke rehabilitation, assessed compensation during exercise performance and annotated the dataset. 

\begin{figure*}[thpb]
      \centering
      \subfloat[\centering E1. \textit{`Brushing Hair'}]{{\includegraphics[width=0.32\textwidth]{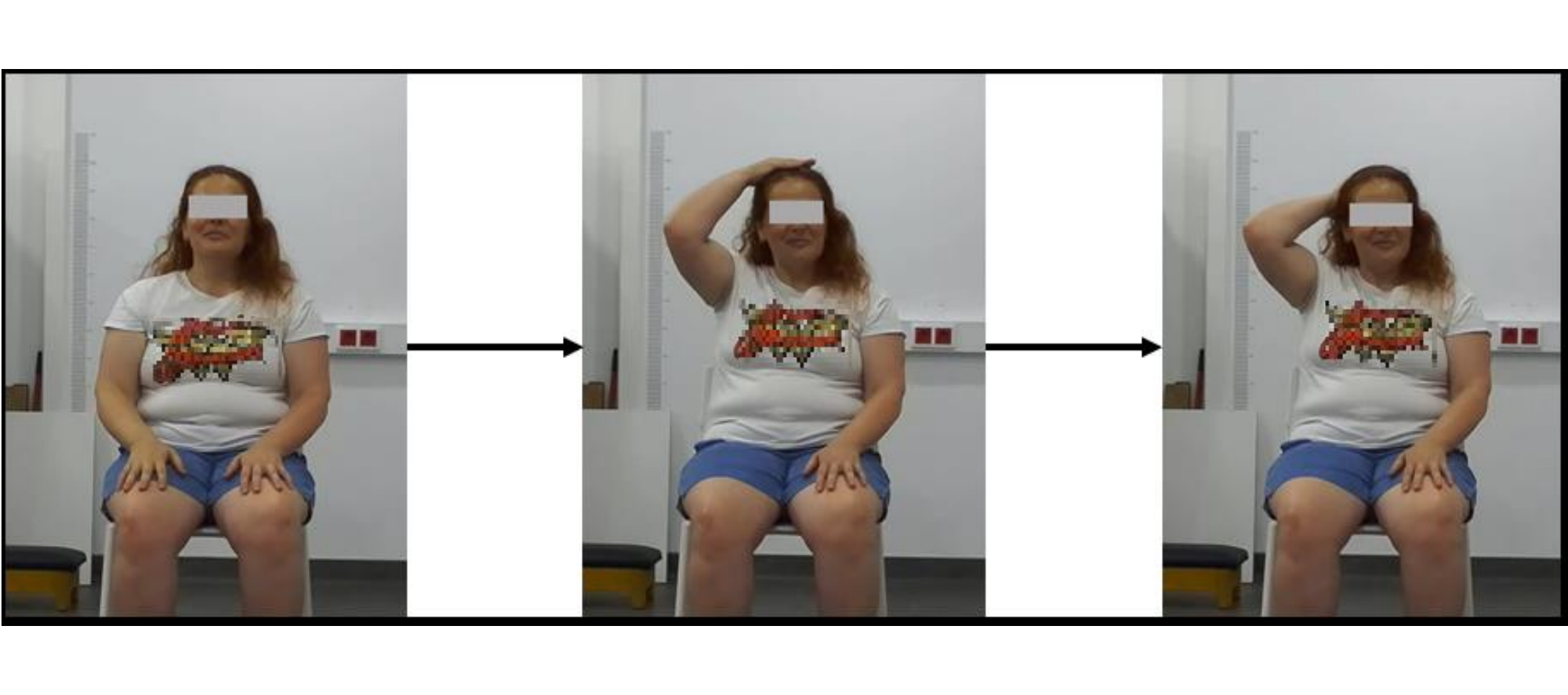} }}%
    \subfloat[\centering E2. \textit{`Brushing Teeth'}]{{\includegraphics[width=0.32\textwidth]{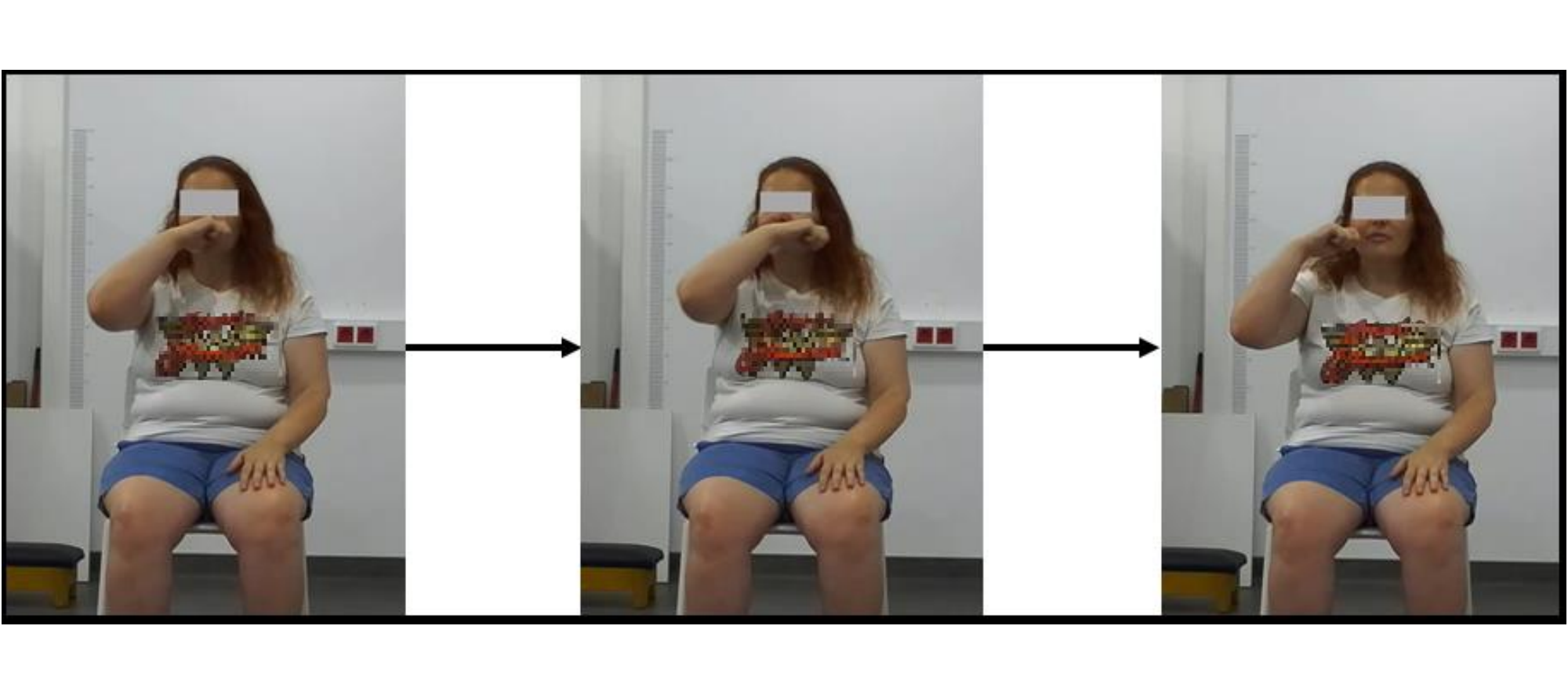} }}%
    \subfloat[\centering E3. \textit{`Wash the Face'}]{{\includegraphics[width=0.32\textwidth]{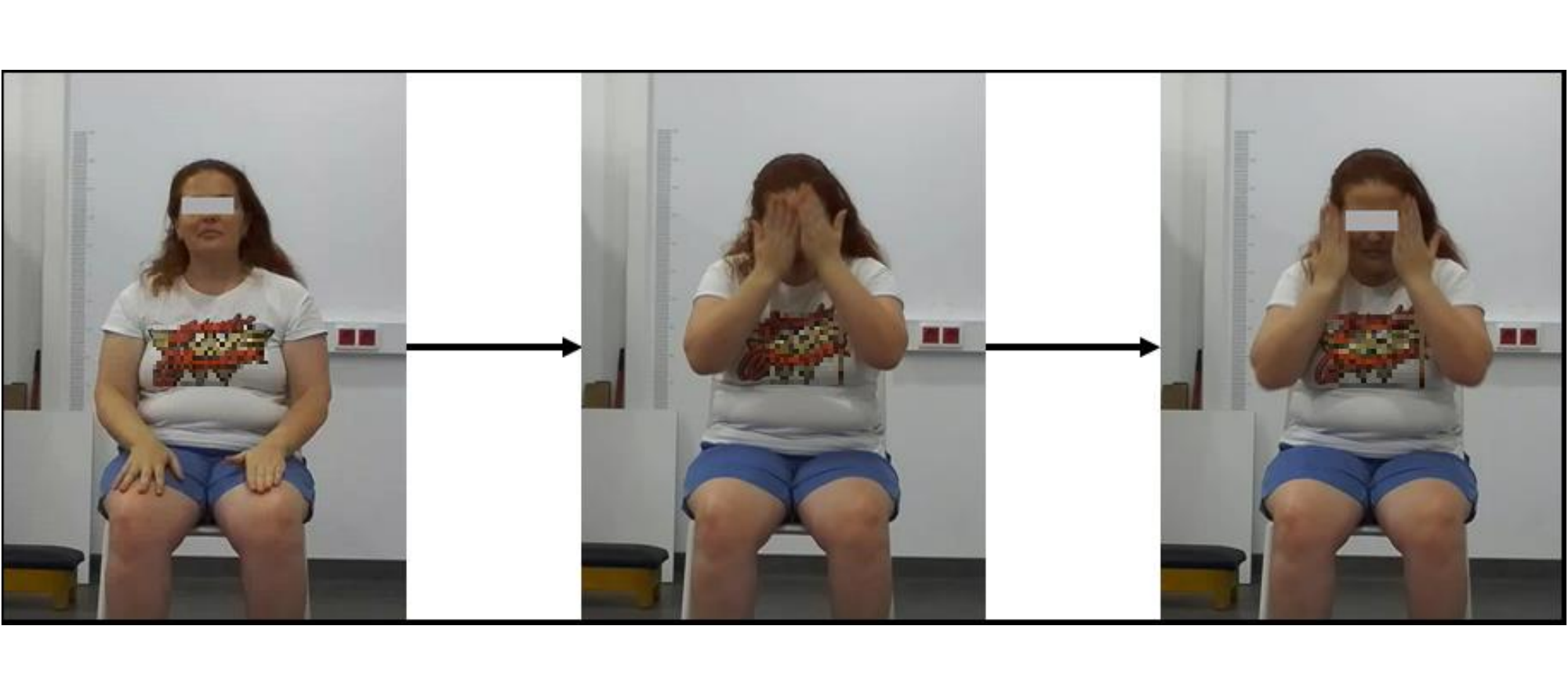} }}%
    \\  
    \subfloat[\centering E4. \textit{`Put on Socks'}]{{\includegraphics[width=0.28\textwidth]{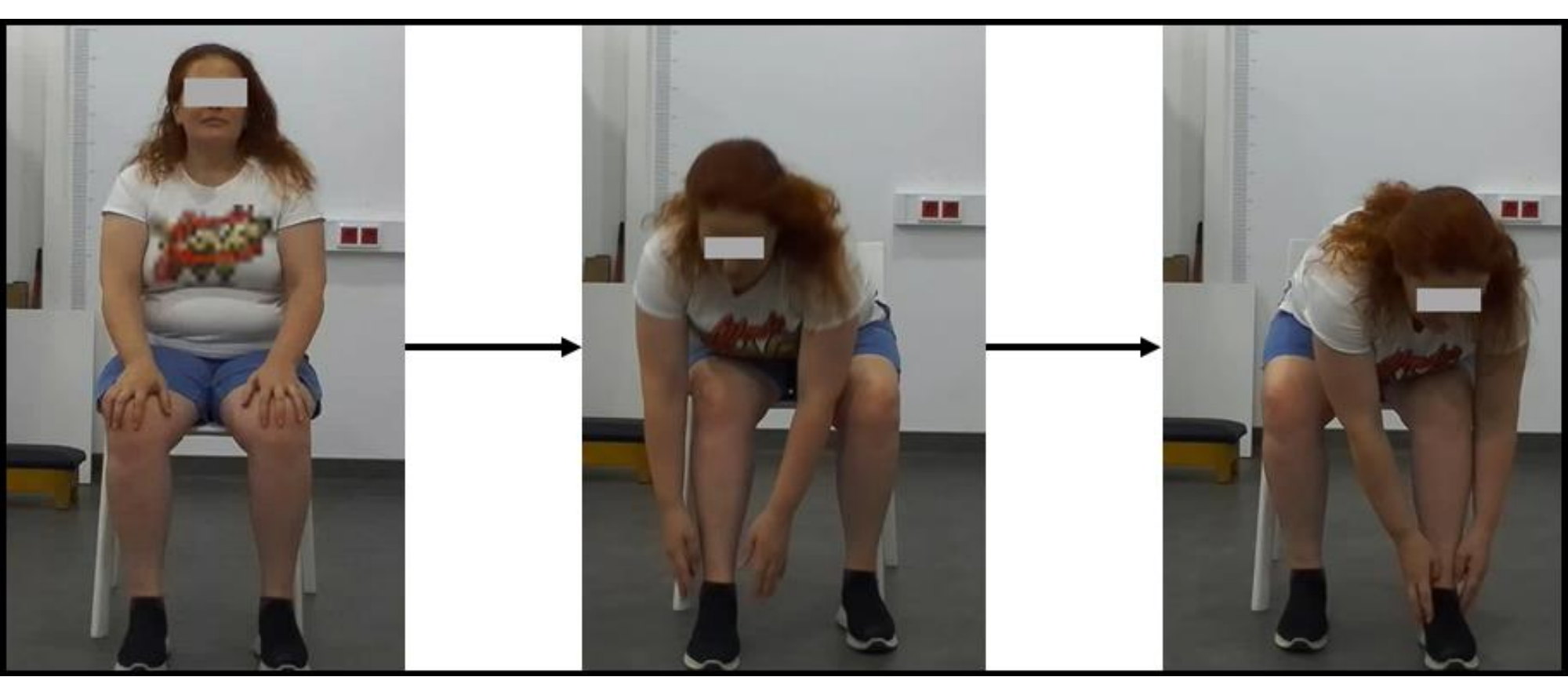} }}%
    \subfloat[\centering E5. \textit{`Hip Flexion'}]{{\includegraphics[width=0.28\textwidth]{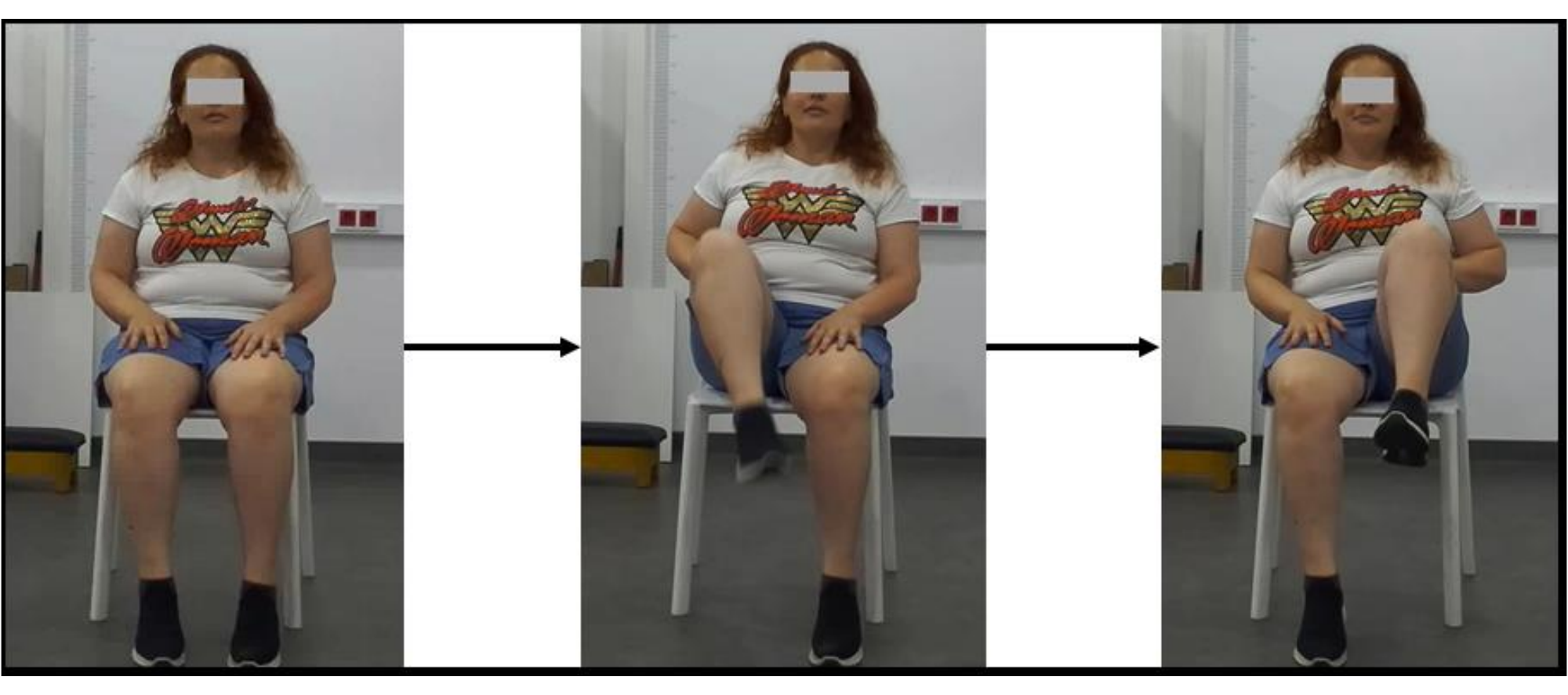} }}
      \caption{ \textit{SERE} functional exercises for rehabilitation.}
      \label{fig:sere}
      \vspace{-0.4cm}
   \end{figure*}

\subsection{Evaluation} 

We evaluate our approach using \textit{Leave-One-Subject-Out} (\textit{LOSO}) cross-validation, where for each run, we fine-tune Model A (box b in Figure~\ref{fig:pipeline}) using data from all post-stroke patients except one, who is held out for testing. From this training set, frame-level pseudo-labels are generated using our pseudo-label selection method, which leverages either Vanilla Gradient or Integrated Gradient saliency maps.

\begin{figure}[H]
    \centering
    \includegraphics[width=0.70\linewidth]{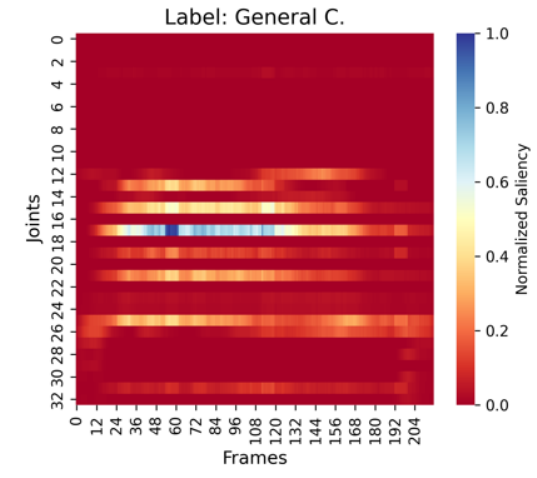}
    \caption{Saliency map generated by the AcT model using the Integrated Gradients (IG) method, for a video of a post-stroke patient performing E1. The video trial is classified as a compensatory movement.}
    \label{fig:saliency_map}
\end{figure}

Figure \ref{fig:saliency_map} shows a saliency map generated by the AcT using the Integrated Gradients (IG) method, trained on a video motion trial of E1 from a post-stroke patient. The model classifies the motion trial as a compensatory motion. On the vertical axis, we see which joints (features) receive the highest relevance scores throughout the trial.

\noindent Since E1 is performed with the left arm, the joints most actively involved are 13, 15, 17, 19, and 21 (Figure \ref{fig:skeleton}). In the saliency map, these joints are highlighted more prominently than in the case of a normal (non-compensatory) motion, indicating a compensatory behavior.

Additionally, for the pseudo-label creation, we inspect the benefit of using a single or dual threshold pseudo-label selection based on specific tradeoffs for False Positives and False Negatives ratios. Table \ref{tab:thresholds} presents the threshold values. These were determined for each model to balance the ratio between False Positives and False Negatives, thereby obtaining pseudo-labels more aligned with the actual distribution of the data and improving the performance of Model B. 

\begin{table}[h!]
\centering
\caption{ Threshold values applied to Vanilla Gradient (VG) and Integrated Gradient (IG) for LSTM, AcT, SkateFormer, and MOMENT.}
\label{tab:thresholds}
\begin{tabular}{|c||cc||cc|}
\hline
\multirow{2}{*}{\textbf{Model}} 
 & \multicolumn{2}{c||}{\textbf{Threshold \#1}} 
 & \multicolumn{2}{c|}{\textbf{Threshold \#2}} \\
\cline{2-5}
 & \textbf{VG} & \textbf{IG} & \textbf{VG} & \textbf{IG} \\
\hline
\textbf{LSTM}        & 4.5 & 2.5 & 4; 6.5  & 2; 3   \\
\textbf{AcT}         & 5.8 & 2.7 & 5.3; 6.3 & 2.2; 3.2 \\
\textbf{SkateFormer} & 3 & 1.5 & 1.5; 2.5 & 0.7; 2   \\
\textbf{MOMENT}      & 2.5 & 3 & 2; 3    & 2; 3.5  \\
\hline
\end{tabular}
\end{table}

Finally, the training data and the pseudo-labels are used to train Model B, a MLP (box d) in Figure \ref{fig:pipeline}), for frame-level assessment. The MLP is tested on the held-out post-stroke patient. This process is conducted 18 times until every post-stroke patient data is used in the test step, assessing all individual motion patterns. We compare the MLP’s predictions to frame-level ground-truth labels to validate the approach.

We use the Area Under the Receiver Operating Characteristic Curve (AUC) to evaluate our approach and experiments. The AUC measures the model's ability to distinguish classes and is unaffected by the pseudo-label selection threshold.

\section{RESULTS AND DISCUSSION}

Models A and B were trained using learning rates (0.001, 0.0001, 0.00001) and dropout probabilities (0.2, 0.3) to optimize generalization and prevent overfitting. The training used the Adam optimizer with a cosine scheduler, binary cross-entropy loss, and ReLU activation, except for the final sigmoid layer. Early stopping prevented overfitting, and batch sizes of 16 and 32 were tested.

\subsection{Video-Level Compensation Assessment - Task Specific vs. Foundation Model}

We evaluate AcT, SkateFormer, and MOMENT performance on video-level compensation assessment against a baseline LSTM. Table \ref{tab:vid_results} summarizes our results. MOMENT yields better results in the task, with an average AUC of $73 \pm 20\%$, outperforming the opponent models with no significant difference ($p>0.05$ using paired t-tests) and the LSTM with statistical significance ($p<0.05$). All models outperformed the baseline LSTM, trained exclusively in \textit{SERE} dataset (AUC = $58 \pm 24\%$), with the AcT and MOMENT revealing statiscally significant difference ($p<0.05$).

Although AcT and SkateFormer are pre-trained in large action recognition datasets, the foundation model for time-series MOMENT provides improved performance in video-level assessment, demonstrating that the knowledge acquired from massive collections of non-task-specific time-series, results in an improved generalization capability to new post-stroke patients with distinctive motion patterns. By leveraging pre-trained knowledge and fine-tuning with the small target dataset, we enhance task accuracy and generalization with less data and reduced training costs.

\begin{table}[ht]
\caption{ Video-level Classification results. Area Under the ROC Curve (AUC) evaluated through \textit{Leave-One-Subject-Out} (\textit{LOSO}) cross-validation strategy across four models. Results are reported as mean $\pm$ standard deviation. All models outperformed the baseline LSTM with MOMENT revealing better performance (pairwise t-tests at $95\%$ significance level).\vspace{-0.2cm}}
\label{tab:vid_results}
\begin{center}
\begin{tabular}{c||c||c||c||c|}
\cline{2-5}
\multicolumn{1}{c||}{}  & \textbf{LSTM} & \textbf{AcT} & \textbf{SkateFormer} & \textbf{MOMENT} \\ \hline
\multicolumn{1}{|c||}{\textbf{AUC}} & 0.58 $\pm$ 0.24 & 0.68 $\pm$ 0.23 & 0.65 $\pm$ 0.23 & \textbf{0.73 $\pm$ 0.20} \\ \hline
\end{tabular}
\end{center}
\end{table}

\subsection{Frame-Level Compensation Assessment}

\begin{table*}[ht]
\caption{Frame-level Classification results. AUC results (mean $\pm$ standard deviation) from \textit{LOSO} cross-validation for Vanilla Gradient (VG) and Integrated Gradient (IG) techniques and video-level assessment models — LSTM, AcT, SkateFormer, and MOMENT. Superscripts indicate statistical comparisons based on paired t-tests at the 95\% confidence level.}
\label{tab:frame_results}
\begin{center}
\begin{tabular}{|c||c||c||c||c||c||c||c||c||c|}
\hline
\multirow{2}{*}{\textbf{$\#$Thr.}} & \multicolumn{2}{c||}{\textbf{LSTM}} & \multicolumn{2}{c||}{\textbf{AcT}} & \multicolumn{2}{c||}{\textbf{SkateFormer}} & \multicolumn{2}{c||}{\textbf{MOMENT}} & \multirow{2}{*}{\begin{tabular}[c]{@{}c@{}} \textbf{Ground} \\ \textbf{Truth} \end{tabular}} \\
\cline{2-9}
 & \textbf{VG} & \textbf{IG} & \textbf{VG} & \textbf{IG} & \textbf{VG} & \textbf{IG} & \textbf{VG} & \textbf{IG} & \\
\hline
\textbf{1 Thr.} 
& 0.64 ± 0.10 & 0.64 ± 0.11 
& 0.53 ± 0.15 & \textbf{0.72 ± 0.11}$^{\dagger}$ 
& 0.58 ± 0.08 & 0.66 ± 0.14 
& 0.51 ± 0.10 & 0.69 ± 0.09$^{\ast}$ 
& \multirow{2}{*}{\begin{tabular}[c]{@{}c@{}} 0.69 \\ ± 0.11 \end{tabular}} \\
\cline{1-9}
\textbf{2 Thr.} 
& 0.64 ± 0.10 & 0.66 ± 0.13 
& 0.55 ± 0.18 & \textbf{0.72 ± 0.10}$^{\dagger}$ 
& 0.64 ± 0.10 & \textbf{0.72 ± 0.12}$^{\ddagger}$ 
& 0.52 ± 0.08 & 0.67 ± 0.12 
& \\
\hline
\end{tabular}
\end{center}
\vspace{-0.4cm}
\begin{flushleft}
\footnotesize
\textbf{Legend:}  
$^{\dagger}$ Not significantly different from Ground Truth ($p>0.05$), MOMENT IG, or AcT 2 Thr.  
$^{\ast}$ MOMENT IG significantly higher than MOMENT VG ($p<0.05$)  
$^{\ddagger}$ SkateFormer IG (2 Thr.) significantly higher than SkateFormer VG (1 Thr.) ($p<0.05$)
\end{flushleft}
\end{table*}

We use the frame-level pseudo-labels to train a MLP for frame-level compensatory motion assessment. We explore whether a Vanilla Gradient (VG) or an Integrated Gradient (IG) technique benefits frame-level classification. In addition, we investigate if applying a single or dual threshold pseudo-label selection approach leads to improved outcomes. Table \ref{tab:frame_results} shows our results for each model, gradient-based technique, and pseudo-label selection approach.

Among all configurations, the AcT model combined with IG techniques achieved the highest frame-level classification performance (AUC = 0.72 ± 0.10), slightly surpassing the performance of the classifier trained with ground-truth labels (AUC = 0.69 ± 0.11). Interestingly, the choice between single and dual thresholding had no impact on AcT’s performance. These results highlight the potential of saliency maps to generate pseudo-labels that are more robust than manual annotations. 

SkateFormer showed improvements with both IG and dual thresholding. The AUC increased from 0.58 ± 0.08 (VG, 1 Thr.) to 0.72 ± 0.12 (IG, 2 Thr.). In contrast, the LSTM model presented limited sensitivity to both gradient type and Selection Method, with results plateauing around 0.64 ± 0.10, and only minor gains observed using IG and dual thresholding (0.66 ± 0.13).

Although MOMENT led in video-level classification, its frame-level performance was comparatively weaker. The model benefited from the use of IG (improving from 0.51 ± 0.10 to 0.69 ± 0.09 with single thresholding), but dual thresholding did not provide further gains.

The AcT IG and the SkateFormer IG, with a dual threshold, surpassed the MLP trained with ground truth labels. Although MOMENT has achieved better performance on the video-level assessment stage, the AcT provided higher quality pseudo-labeling, enabling a performance on the frame-level assessment even better than using frame-level ground truth labels. Working on top of accurate video-level predictions for frame-level pseudo-label automatic creation by thresholding gradient-based saliency maps can result in more precise labeling, leading to improved frame-level classification results compared with a model trained on ground truth labeling. Ground truth data labeling can be noisy as human annotators are prone to mistakes and, usually, their labeling methodology is based on subjective data evaluation. Labeling is based on therapists' subjective evaluation of post-stroke patients' motions, in which the level of agreement might be low \cite{lee2019learning}, and depends on professionals' experience. Additionally, detecting compensatory motion boundaries is highly challenging \cite{lee2020towards}, leading to labeling inaccuracies.

\section{CONCLUSIONS}

In this work, we introduced a framework for real-time assessment of compensatory motion in stroke rehabilitation exercises using video-level annotations to train frame-level classifiers. By exploiting gradient-based saliency maps and a pseudo-label selection method, we addressed the challenge of obtaining frame-level labels from video-level annotations, significantly reducing manual labeling efforts. We leverage pre-trained task-specific models and a foundation model to enhance generalization to new patients.

Our experiments demonstrated that the foundation model MOMENT achieves superior performance in video-level classification (AUC = $73 \pm 20\%$), outperforming task-specific models and the LSTM baseline (AUC = $58 \pm 24\%$). For frame-level assessment, the Action Transformer (AcT) model, combined with Integrated Gradients (IG), delivered the best results (AUC = $72 \pm 10\%$), even surpassing a model trained on ground-truth frame-level labels (AUC = $69 \pm 11\%$). These results suggest that pseudo-labels generated from accurate video-level predictions and gradient-based saliency maps can be more reliable than manual annotations, which are susceptible to human bias. Our approach reduces the efforts required for frame-level annotation, facilitating the training of new models while enhancing generalization to new patients, which is crucial in rehabilitation scenarios where individual motor patterns vary significantly.

To further improve the robustness of our approach, future work will focus on enhancing model generalization across different rehabilitation exercises. Additionally, we plan to explore the application of this technique to other datasets beyond the stroke rehabilitation domain, such as general action recognition benchmarks. This will help assess the method's ability to generalize across different types of movement patterns and tasks, providing insights into its broader applicability in human motion analysis.

\section*{ACKNOWLEDGMENTS}

The authors would like to thank their colleagues from the Robot Vision Laboratory (VisLab), Laboratory of Robotics and Engineering Systems (LARSyS), Instituto Superior Técnico, and Mononito Goswami of the CMU Auton Lab, Carnegie Mellon University, for their insightful discussions, technical support, and valuable feedback throughout this research.



\bibliographystyle{IEEEtran}
\bibliography{IEEEabrv,mybibfile}

\end{document}